\documentclass[12pt,twoside]{article}
\usepackage{amssymb}
\usepackage{amsmath}
\usepackage[makeroom]{cancel}
\usepackage{latexsym}
\usepackage{longtable}
\usepackage{epsfig}
\usepackage{dsfont}
\usepackage{graphicx,bbm,psfrag}
\graphicspath{{images/}}
\usepackage[ascii]{inputenc}
\numberwithin{equation}{section}

\begin{document}
\vspace*{2cm}
\begin{center}
 {\LARGE Generalized Gibbs Ensembles of the Classical \bigskip\\Toda Chain}\bigskip\bigskip
 \end{center}
\begin{center}
{\large Herbert Spohn}
 \end{center}
 \begin{center}
Zentrum Mathematik and Physik Department, TUM,\\
Boltzmannstra{\ss}e 3, 85747 Garching, Germany,
 \tt{spohn@tum.de}
 \end{center}
%\maketitle
\vspace*{3cm}
\noindent Version 3 of this preprint is published online by Journal of Statistical Physics. Unfortunately, 
 on the right hand side of the two equations in \eqref{2.19} a factor $2$ is missing. To properly correct, factors of $2$ dis- and reappear in various formulas throughout Sections 2 and 3.  In this version the text, including references, remains untouched, only the factors of 2 are properly taken care of.
\vspace*{1cm}\\
\textbf{Abstract}. The Toda chain is the prime example of a classical integrable system with strictly local conservation laws.
Relying on the Dumitriu-Edelman matrix model, we obtain the generalized free energy of the Toda chain and thereby establish a mapping  
to the one-dimensional log-gas with an interaction strength of order $1/N$. The (deterministic) local density of states of the Lax matrix
is identified as the object, which should evolve according to generalized hydrodynamics.\\\\\\
\textit{It is my great pleasure to dedicate this article to Joel Lebowitz as teacher, as guide to yet unexplored scientific territories, and with gratitude for a lasting friendship. }

\vspace*{2cm}
\begin{flushright} 25.11.2019
\end{flushright}

%%%%%%%%%%%%%%%%%%%%%%%%%%%%
\newpage
 \section{Introduction}
\label{sec1}
\setcounter{equation}{0} 
Fluid hydrodynamics is based on the notion of local equilibrium. If in the initial state the thermodynamic parameters are slowly varying in space, then 
they will vary also slowly in time. As a consequence the local equilibrium parameters satisfy the Euler equations which is a closed set of hyperbolic conservation laws.  The structure of the 
Euler equations is universal, while the specific fluid properties enter through a model-dependent  pressure. The central cornerstone of hydrodynamic theory is 
the validity of  \textit{local} microscopic conservation laws. The underlying microscopic dynamics could be either classical, or quantum, or stochastic. If the Navier-Stokes corrections are included, then the time range  accessible to hydrodynamics encompasses even the approach to thermal equilibrium.

Over the recent years the hydrodynamic approach has been most successfully extended to integrable quantum chains. The main novel feature is an infinite
number of conservation laws. Thus thermal equilibrium must be replaced by a generalized Gibbs ensemble (GGE), which depends on an infinite set of chemical potentials.
However, to  even write down the macroscopic conservation laws one has to compute the GGE average of the conserved fields and their currents. For the field averages  
a suitable extension of the Bethe ansatz usually suffices. The real difficulty consists then in finding out about the average currents, which for interacting systems cannot be written as derivatives of  the GGE free energy. The generic structure of the currents was first derived in \cite{CDY16,
PNCBF17}, leading to a coupled set of equations called generalized hydrodynamics (GHD). Fully worked out examples have been accomplished, as the quantum sinh-Gordon model \cite{CDY16},
the XXZ spin chain \cite{PNCBF17}, the spin $\tfrac{1}{2}$ Hubbard model \cite{IN17}, and the Lieb-Liniger model \cite{CDY16} with a recent experimental confirmation  \cite{SBDD18}. 

With this background, one might wonder about classical integrable systems.
The still instructive blue-print is a fluid of  hard rods in one dimension \cite{BDS83,S91}, for which domain wall initial states have been studied
\cite{DS17a} and even the Navier-Stokes correction has been proved \cite{BS97}. In the context of the Korteweg-de-Vries equation, Zakharov 
\cite{Z09} proposed a hydrodynamic type theory. We refer to \cite{CDE16} for more details. For the relativistic sinh-Gordon model, also an integrable classical 
field theory, GHD is derived and studied  in \cite{BDWY18}. Besides their intrinsic interest, classical models are more accessible to 
 molecular dynamics simulations. A system  size  of $10^4$ particles is standard, while quantum DMRG simulations would not go much beyond $10^2$
 spins. 
Accurate simulations are already available for the Toda lattice \cite{KD16}, the Faddeev-Takhtajan spin chain \cite{DD18}, and a discrete integrable version 
of the sinh-Gordon model \cite{BDWY18}. With classical models one has a better chance to test the range of validity of GHD. 

In this note I will consider the classical Toda chain. My main novel observation is to connect the Dumitriu-Edelman matrix model with the Toda Lax matrix in thermal equilibrium.  The generalized Gibbs free energy of the Toda chain turns out to be related to the $\beta$-ensembles of random matrix theory in the mean-field regime, for which the 
$\beta$-parameter scales as $1/N$. Thereby an exact variational formula is obtained for the density of states of the Lax matrix, when its matrix elements are distributed according to some GGE.
\newpage
%%%%%%%%%%%%%%%%%%%%%%%%%%%%%%%%%%%%%%%%%%%%%%%%%%%%%%%%%%%%%%%%%%%%%%
\section{Toda lattice, local conservation laws, and GGE}
\label{sec2}
\setcounter{equation}{0} 

Following the notation of M. Toda \cite{T89}, Section 3, the hamiltonian of the Toda chain on sites $j = 1,...,N$ is given by 
\begin{equation}\label{2.1}
H = \sum_{j =1}^N\big( \tfrac{1}{2}p_j^2 + \mathrm{e}^{-r_j}\big),\quad r_j = q_{j+1} - q_j,
\end{equation}
where periodic boundary conditions, $q_{N+j} = q_j +\ell$, are assumed. The positional increments, $r_j$, are the stretches. They tell us about
the elongation/compression of the chain and can be of either sign. The equations of motion read
 \begin{equation}\label{2.1a}
\frac{d}{dt}q_j = p_j, \qquad \frac{d}{dt}p_j =  \mathrm{e}^{-r_{j-1}}  -\mathrm{e}^{-r_j},
\end{equation}
and are viewed as a discrete nonlinear wave equation, just as the discrete nonlinear Schr\"{o}din\-ger equation.
The underlying lattice of labels is $\mathbb{Z}$, resp. some bounded interval of $\mathbb{Z}$. 
Alternatively, one can interpret \eqref{2.1a} as labelled point particles moving on the real line, $\mathbb{R}$, see \cite{D19} for more details.
We define
\begin{equation}\label{2.2}
a_j = \mathrm{e}^{-r_j/2},\quad b_j =  p_j.
\end{equation} 
Then the finite $N$ Lax matrix \cite{F74} is the tridiagonal real symmetric matrix with
\begin{equation}\label{2.3}
(L_N)_{j,j} = b_j, \quad (L_N)_{j,j+1} = (L_N)_{j+1,j}= a_j, 
\end{equation} 
  $(L_N)_{1,N} = (L_N)_{N, 1} = a_N$ because of periodic boundary conditions, and $(L_N)_{i,j}=0$ otherwise. 
There is also the antisymmetric $B_N$ matrix, $B_N = - (B_N)^\mathrm{T}$, with
\begin{equation}\label{2.4}
(B_N)_{j,j} = 0, \quad (B_N)_{j,j+1} = -  (B_N)_{j+1,j}= -a_j,    
\end{equation} 
 $(B_N)_{1,N} = - (B_N)_{N, 1} = a_N$ because of periodic boundary conditions, and $(B_N)_{i,j}=0$ otherwise. 
 Then, under the Toda dynamics,
 \begin{equation}\label{2.4a}
\frac{d}{dt} L _N= [\tfrac{1}{2}B_N,L_N].
\end{equation} 
 Since $B_N$ is antisymmetric, the eigenvalues of $L_N$ are time-independent and the 
locally conserved fields $Q^{[n],N}$, $n = 1,..., N$, of the Toda lattice can be written as  
\begin{equation}\label{2.5} 
Q^{[n],N} = \mathrm{tr}\big[(L_N)^n\big] = \sum_{j=1}^N ((L_N)^n)_{j,j}. 
\end{equation} 

Switching to the infinite lattice. As a generic result, the dynamics of the chain is well defined for initial conditions
$\{q_j,p_j,j\in\mathbb{Z}\}$, which increase at infinity slower than exponential, see \cite{butta} for more precise statements.
Let us introduce the two-sided infinite Lax matrix, $L$,  by 
\begin{equation}\label{2.6}
L_{j,j} = b_j, \quad L_{j,j+1} = L_{j+1,j}= a_j, 
\end{equation} 
and $L_{i,j}=0$ otherwise, $i,j\in\mathbb{Z}$. From \eqref{2.5} we infer that the conserved fields have a density  given by 
\begin{equation}\label{2.7}
Q_{j}^{[n]} = (L^n)_{j,j}
\end{equation} 
at site $j$. Note that $Q_{j}^{[n]}$ depends at most on the variables in $[j-n,...,j+n]$. Explicit expressions for $n=1,...,5$ are listed in \cite{H74}. 
In particular, for momentum and energy, 
\begin{equation}\label{2.8}
Q_{j}^{[1]} = p_j,\quad \tfrac{1}{2}Q_{j}^{[2]} = \tfrac{1}{2}\big(p_j^2 +   \mathrm{e}^{-r_{j-1}} + \mathrm{e}^{-r_j}\big).
\end{equation} 

Since there is an explicit expression for $Q_{j}^{[n]}$, one can determine the currents defined through
\begin{equation}\label{2.9}
\frac{d}{dt}Q_{j}^{[n]} = J_{j}^{[n]} - J_{j+1}^{[n]}.
\end{equation} 
Using that 
\begin{equation}\label{2.10}
\frac{d}{dt} L = [\tfrac{1}{2}B,L],
\end{equation} 
with $B$ the two-sided infinite version of $B_N$, one obtains
 \begin{equation}\label{2.11}
\frac{d}{dt} L^n = [\tfrac{1}{2}B,L^n]
\end{equation}
and 
\begin{equation}\label{2.12}
\frac{d}{dt} Q_{j}^{[n]} = (BL^n - L^nB)_{j,j} = - a_j (L^n)_{j+1,j} + a_{j-1}(L^n)_{j,j-1}.
\end{equation} 
Since $L$ is symmetric, the second term is the left shift of the first term and hence one local version of the current is
\begin{equation}\label{2.13}
J_{j}^{[n]} = \tfrac{1}{2}(L^nL^\mathrm{off})_{j,j}
\end{equation} 
with $L^\mathrm{off}$ denoting the off-diagonal part of $L$. The total current at finite volume $N$  can be written as 
 \begin{equation}\label{2.13a}
J^{[n],N} = \tfrac{1}{2}\mathrm{tr}\big[(L_N)^nL_N^\mathrm{off}\big].
\end{equation} 

We turn to the generalized Gibbs ensembles (GGE) of the finite $N$ Toda chain, our main goal being their infinite volume free energy.
Since the Toda lattice has many conserved fields, the canonical prescription is the introduce a chemical potential, $\mu_n$, for each field, 
i.e. to consider
\begin{equation}\label{2.14}
Z_N^{-1} \exp\Big[\sum_{n=0}^\kappa \mu_n Q^{[n]}\Big],
\end{equation}
where $\kappa >0$ is assumed to be even and $\mu_\kappa <0$. 
 If the conventional picture of thermalization for isolated systems still applies to integrable models, one would expect that in the long time limit the statistics of \textit{local} observables would be provided by \eqref{2.14}. Of course, the chemical potentials have to be chosen in such a way that the expectation values of the conserved fields in the initial state are matched. It will be convenient to have the short hand
\begin{equation}\label{2.14a}
V(x) = -\sum_{n=0}^\kappa\mu_n x^n.
\end{equation}
Then the GGE can be written as 
 \begin{equation}\label{2.14b}
Z_N^{-1} \exp\big(- \mathrm{tr}[V(L_N)]\big).
\end{equation}

Physically our expression for the GGE misses the fact that the compression of the Toda lattice will also change on the hydrodynamic scale. In case of a generic anharmonic chain, an external  pressure is enforced by, for example, fixing the left end and
pushing/pulling the right end. From the equivalence of ensembles, see the discussion in  \cite{BerOll,S14}, it follows that the exponential of \eqref{2.14}
has to be complemented by the stretch field
\begin{equation}\label{2.15}  Q^{[\mathrm{s}],N} = \sum_{j =1}^N r_j
\end{equation} 
with dual parameter $P$, which controls the pressure in the chain. Only for $P> 0$ one has a well-defined GGE. For $P =0$, even worse for $P<0$,  the chain would simply fall apart.
From the equations of motion \eqref{2.1a},
obviously $r_j$ is locally conserved with corresponding current $-p_j$. Hence we complete the list of conservation laws by
\begin{equation}\label{2.16}
  Q^{\mathrm{[s]}}_{j} = r_j, \quad J^{\mathrm{[s]}}_{j} = -p_j.
\end{equation} 
The finite volume GGE is then
\begin{equation}\label{2.17}
\mu_N^{(V,P)}(\mathrm{d}^Nr \mathrm{d}^Np) = (Z_{\mathrm{toda},N})^{-1}\mathrm{e}^{-\mathrm{tr}[V(L_N)]}\prod_{j=1}^N \mathrm{e}^{ -Pr_j}\mathrm{d}r_j\mathrm{d}p_j, \quad P>0.
\end{equation} 
Since  $\mu_\kappa <0$, there are constants $c_0, c_2$, $c_2 > 0$, such that $V(x) \geq c_0 + c_2x^2$ and hence
\begin{equation}\label{2.17c}
\mathrm{tr}[V(L_N)] \geq c_0N + c_2 \mathrm{tr}[(L_N)^2] ,
\end{equation} 
which ensures a finite partition function.

Thermal equilibrium corresponds to the choice $V(x) = \tfrac{1}{2}\beta x^2$, $\beta >0$ the inverse temperature. [Below, $\beta$ will also be used as parameter of the log-gas. Given the context, no ambiguity should arise.] The physical pressure, $\tilde{P}$, is equated with the thermally averaged force between a nearest neighbor pair of particles, which then yields $P = \beta \tilde{P}$.  For convenience $P$ is still  called pressure. 
In thermal equilibrium the average stretch is computed as
\begin{equation}\label{2.17b}
\langle r_0\rangle_{\tilde{P}} = \log \beta - \frac{\Gamma'(\beta \tilde{P})}{\Gamma(\beta \tilde{P})}.
\end{equation} 
 Hence there is a critical $\tilde{P}_\mathrm{c}$ such that  $\langle r_0\rangle_{\tilde{P}_\mathrm{c}} = 0$. 
 For $\tilde{P} < \tilde{P}_\mathrm{c}$ the positions are predominantly ordered as $ q_j < q_{j+1}$ and $\langle r_0\rangle_{\tilde{P}}\to \infty$
  as $\tilde{P} \to 0$. On the other hand, for $\tilde{P} > \tilde{P}_\mathrm{c}$ the positions  are predominantly 
reversely ordered and $ \langle r_0\rangle_{\tilde{P}} \simeq -\log \tilde{P}$. 

One might wonder, whether one could soften the somewhat stringent assumption \eqref{2.14a}. In fact, for later considerations only a confining potential  $V$
with some smoothness will be  required. Not to interrupt the flow of the argument we collect some technical remarks in the Appendix. 
 
We split as $V(x) = \tfrac{1}{2} x^2 + \tilde{V}(x)$ to obtain
\begin{equation}\label{2.17a}
Z_{\mathrm{toda},N} = Z_{L,N}(Z_{L,N})^{-1}\int_{\mathbb{R}^{2N}}\prod_{j=1}^N \mathrm{d}r_j\mathrm{d}p_j
\mathrm{e}^{ -Pr_j}\mathrm{e}^{-\frac{1}{2}\mathrm{tr}[(L_N)^2]} \mathrm{e}^{-\mathrm{tr}[\tilde{V}(L_N)]}.
\end{equation}
By an explicit computation
\begin{equation}\label{2.18}
Z_{L,N} = \int_{\mathbb{R}^{2N}}\prod_{j=1}^N \mathrm{d}r_j\mathrm{d}p_j \mathrm{e}^{ -Pr_j}
\mathrm{e}^{-\frac{1}{2}\mathrm{tr}[(L_N)^2]}= \big(\sqrt{2\pi}\Gamma(P)\big)^N.
\end{equation} 
Thus in \eqref{2.17a} we average $\exp(-\mathrm{tr}[\tilde{V}(L_N)])$ over a normalized reference measure, its expectation being denoted by $\mathbb{E}_{L_N}(\cdot)$. Under this measure $L_N$  becomes a symmetric \textit{random} Jacobi matrix. 
Let $\xi_\mathrm{G}$ be a unit Gaussian random variable, probability density function
$(2\pi)^{-\frac{1}{2}}\exp\big(-x^2/2\big)$, and let $\chi_k$ a chi-distributed random variable with parameter $k>0$, probability density function
$\big(2^{(k/2) - 1}\Gamma(k/2)\big)^{-1}x^{k-1}\exp\big(-x^2/2\big)$ for $x>0$ and zero otherwise. Then $\{(L_N)_{j,j}, j =1,...,N\}$ and $\{(L_N)_{j,j+1},j =1,...,N\}$ are two independent families  of i.i.d random variables,
where in distribution
\begin{equation}\label{2.19}
L_{1,1} =  \xi_\mathrm{G}, \quad L_{1,2} = \tfrac{1}{\sqrt{2}}\chi_{2P}.
\end{equation} 
The free energy of the Toda chain is now expressed as
\begin{equation}\label{2.20}
F_\mathrm{toda}=  -\lim_{N \to \infty} \tfrac{1}{N} \log Z_{\mathrm{toda},N} = - \log(\sqrt{2\pi} \Gamma(P))  +  F_\mathrm{L},
\end{equation}
where
\begin{equation}\label{2.21}
F_\mathrm{L} = -\lim_{N \to \infty} \tfrac{1}{N}\log\mathbb{E}_{L_N}\big( \mathrm{e}^{-\mathrm{tr}[
\tilde{V}(L_{N})]}\big).
\end{equation}
The Toda free energy depends on the potential $V$ and the pressure $P$, and likewise $F_\mathrm{L}$. Using the standard H\"{o}lder inequality for the partition function, one obtains  
for $0 \leq \eta \leq 1$
\begin{equation}\label{2.22}
F_\mathrm{toda}((1-\eta) V_0 + \eta V_1,(1-\eta) P_0 + \eta P_1) \leq   
(1-\eta)F_\mathrm{toda}(V_0, P_0) + \eta F_\mathrm{toda}(V_1,P_1).
\end{equation}
Thus as a function of the intensive parameters $F_\mathrm{toda}$ is convex down, while being convex up as function of the extensive fields.
In particular, for fixed $V$,  $F_\mathrm{toda}(P)$ is convex down and $\partial_P F_\mathrm{toda}(P) =  \langle r_0\rangle_{P}$ is decreasing,
consistent with the explicit case \eqref{2.17b}. To keep the notation slim,
the dependence on $V,P$ is not displayed, except where needed from the context.

The expression \eqref{2.21} looks like an unsurmountable obstacle. To my own surprise random matrix theory turns out to be the supportive tool.

%%%%%%%%%%%%%%%%%%%%%%%%%%%%%%%%%%%%%%%%%%%%%%%%%%%%
\section{Random matrix model, free energy}
\label{sec3}
\setcounter{equation}{0} 

Dumitriu and Edelman \cite{DE02} introduced a Jacobi matrix model which has the same eigenvalue distribution as the 
$\beta$-ensemble from random matrix theory \cite{RM}.  Their $N \times N$ matrix model is
\begin{equation}\label{3.1}
(T_{N,\beta})_{j,j} =  \xi_{\mathrm{G},j}, \quad (T_{N,\beta})_{j,j+1} = (T_{N,\beta}) _{j+1,j} = \tfrac{1}{\sqrt{2}}\chi_{(N-j)\beta}, \quad \beta > 0,
\end{equation} 
with $(T_{N,\beta})_{i,j} = 0$ otherwise, $i,j = 1,...,N$. The random variables $\{\xi_{\mathrm{G},j},\chi_{(N-j)\beta}, j = 1,...,N\}$ are independent. The average with respect to $T_{N,\beta}$  is denoted by
$\mathbb{E}_{T_{N,\beta}}(\cdot) $.\bigskip\\
\textbf{Proposition 3.1}. \cite{DE02} \textit{Let $\{\lambda_1,..., \lambda_N\}$ be the eigenvalues of $T_{N,\beta}$. Then their 
joint probability density function is given by the $\beta$-ensemble of random matrix theory, namely
\begin{equation}\label{3.2}
(Z_{N,\beta})^{-1} \prod_{1\leq i < j \leq N}\big|\lambda_i - \lambda_j\big|^{\beta} \prod_{j=1}^N \mathrm{e}^{-\frac{1}{2}\lambda_j^2}\mathrm{d}\lambda_j 
\end{equation}
with partition function}
\begin{equation}\label{3.3}
Z_{N,\beta} = (2\pi)^{N/2}\Gamma(1 +\tfrac{1}{2}\beta)^{-N} \prod_{j=1}^N\Gamma(1+ \tfrac{1}{2}\beta j).
\end{equation} \\
To establish the connection with the Lax matrix we set $\beta = 2P/N$.  Then  locally close to $\lfloor (1-u)N\rfloor$,  $0 \leq u \leq 1$, the  matrix $T_{N,2P/N}$ will look like $L_{u P }$. Since the  parameter of the $\chi$ random variables changes on scale $1/N$, the corresponding free  energies merely add up, see Appendix.
Thus, setting
 \begin{equation}\label{3.4}
F_\mathrm{T}(P) = -\lim_{N\to \infty}\tfrac{1}{N}\log\mathbb{E}_{T_{N,2P/N}}\big( \mathrm{e}^{-\mathrm{tr}[ 
\tilde{V}(T_{N,2P/N})]}\big),
\end{equation}
one concludes that
\begin{equation}\label{3.5} 
F_\mathrm{T}(P) = \int_0^1 \mathrm{d}u F_\mathrm{L}(uP).
\end{equation}

Relying on Proposition 3.1,
\begin{eqnarray}\label{3.6} 
&&\hspace{-30pt}\mathbb{E}_{T_{N,2P/N}}\big( \mathrm{e}^{-\mathrm{tr}[
\tilde{V}(T_{N,2P/N})]}\big)\nonumber\\ 
&&= (Z_{N,2P/N})^{-1} \int_{\mathbb{R}^N}\prod_{j=1}^N \mathrm{d}\lambda_j \exp\Big[- \sum_{j=1}^N V(\lambda_j) 
+\frac{1}{N}\sum_{i \neq j=1}^N  P\log| \lambda_i - \lambda_j| \Big],
\end{eqnarray}
where the quadratic term in \eqref{3.2} recombines with $\tilde{V}$ to arrive at $V$. Thus the large $N$ limit of the properly scaled Dumitriu-Edelman matrix is reduced to the mean-field limit for a conventional statistical mechanics model. 
Its central object is the empirical density
\begin{equation}\label{3.6a}
\rho_N(\mathrm{d}x) = \frac{1}{N} \sum_{j=1}^N \delta (x- \lambda_j)\mathrm{d}x.
\end{equation}
As $N \to \infty$ the sequence $\rho_N(\mathrm{d}x)$ converges to the  almost sure limit, $\rho^*(x) \mathrm{d}x$, which is determined by a variational formula.
For a detailed and instructive discussion we refer to \cite{R15}, Section 2.3. Note that in the following discussion $V$ remains fixed.

More precisely one defines the free energy functional
\begin{equation}\label{3.7}
 \mathcal{F}_P^\mathrm{MF}(\rho) =  \int _\mathbb{R}\mathrm{d}x \rho(x) V(x)      -P \int _\mathbb{R}\mathrm{d}x\int _\mathbb{R}\mathrm{d}y  \log|x-y|  \rho(x)\rho(y) + 
\int _\mathbb{R}\mathrm{d}x \rho(x) \log \rho(x).
\end{equation} 
The first two terms are the obvious continuum limit of the corresponding expressions  in \eqref{3.6}. The entropy term results from the product reference measure.
$\rho^*$ is then the minimizer of $ \mathcal{F}_P^\mathrm{MF}$ under the constraint 
\begin{equation}\label{3.8}
\rho(x) \geq 0,\quad  \int _\mathbb{R}\mathrm{d}x \rho(x) =1 
\end{equation}
and the limiting free energy associated to  \eqref{3.6} is given by $\mathcal{F}_P^\mathrm{MF}(\rho^*)$.
Since $\mathcal{F}_P^\mathrm{MF}$ is convex, the minimizer is unique.
Then 
\begin{equation}\label{3.9} 
F_\mathrm{T}(P) = \big(\log\sqrt{2\pi}  + \int_0^1 \mathrm{d}u \log \Gamma(1+uP) \big) + \mathcal{F}_P^\mathrm{MF}(\rho^*).
\end{equation}
 Hence
\begin{equation}\label{3.10} 
 \int_0^1 \mathrm{d}u F_\mathrm{toda}(uP) =  \mathcal{F}^\mathrm{MF}_P(\rho^*) +\log P  -1.
\end{equation}
Changing variables and differentiating with respect to $P$ yields
 \begin{equation}\label{3.11} 
 F_\mathrm{toda}(P) =  \partial_P\big( P\mathcal{F}_P^\mathrm{MF}(\rho^*) \big) +\log P . 
\end{equation}

$P$ appears only in the strength of the quadratic interaction term. Thus it turns out to be more convenient to introduce $\varrho = P \rho$ and
the modified free energy functional 
\begin{equation}\label{3.12}
\mathcal{F}(\varrho) =  \int _\mathbb{R}\mathrm{d}x \varrho(x) V(x)      - \int _\mathbb{R}\mathrm{d}x\int _\mathbb{R}\mathrm{d}y   \log|x-y|\varrho(x) \varrho(y) + 
\int _\mathbb{R}\mathrm{d}x \varrho(x) \log \varrho(x).
\end{equation} 
Then $P \mathcal{F}_P^\mathrm{MF}(P^{-1}\varrho)=  \mathcal{F}(\varrho) -P\log P$.
The $P$-dependence has now been moved to the constraint as
\begin{equation}\label{3.13}
\varrho(x) \geq 0,\quad  \int _\mathbb{R}\mathrm{d}x \varrho(x) =P. 
\end{equation}
We denote the minimizer of $\mathcal{F}$ by $\varrho^*(P)$. Then 
 \begin{equation}\label{3.14} 
 F_\mathrm{toda}(P) =  \partial_P \mathcal{F}(\varrho^*(P)) - 1.
 \end{equation}
 
 To remove the constraint one introduces the Lagrange multiplier $\lambda$ and sets
  \begin{equation}\label{3.15} 
 \mathcal{F}_\lambda(\varrho) =  \mathcal{F}(\varrho) - (\lambda + 1) \int _\mathbb{R}\mathrm{d}x \varrho(x).
 \end{equation}
A minimizer of  $\mathcal{F}_\lambda(\varrho)$ is denoted by  $\varrho_\lambda^*$ as determined by the solution of the Euler-Lagrange equation
\begin{equation}\label{3.16} 
  V(x) -  2 \int_\mathbb{R} \mathrm{d}y  \log|x-y| \varrho_\lambda^*(y) +\log \varrho_\lambda^*(x)  - \lambda = 0.
 \end{equation}
$\lambda$ has to be adjusted such that
\begin{equation}\label{3.17} 
 P =  \int _\mathbb{R}\mathrm{d}x  \varrho_\lambda^*(x).
 \end{equation}
 To obtain the Toda free energy, we differentiate as
 \begin{eqnarray}\label{3.18} 
 && \hspace{-40pt} \partial_P \mathcal{F}(\varrho^*(P)) =
  \int_\mathbb{R} \mathrm{d}x \partial_P\varrho^*(x,P) V(x)      - 2 \int_\mathbb{R} \mathrm{d}x\int_\mathbb{R} \mathrm{d}y \log|x-y| \partial_P\varrho^*(x,P)\varrho^*(y,P)\nonumber\\
 && \hspace{40pt} 
 + \int_\mathbb{R} \mathrm{d}x \partial_P\varrho^*(x,P)  \log \varrho^*(x,P) +1 .
  \end{eqnarray}
 Thus integrating \eqref{3.16} against  $\partial_P\varrho^*(P)$ one obtains 
 \begin{equation}\label{3.19} 
 \partial_P \mathcal{F}(\varrho^*(P)) = \lambda +1.
\end{equation}
 From \eqref{3.14} it then follows that 
 \begin{equation}\label{3.20} 
  F_\mathrm{toda}(P) =  \lambda(P).
 \end{equation}
 
As mentioned already, $ F_\mathrm{toda}(P)$, hence $\lambda(P)$, is convex down and $\lambda'(P) = \langle r_0 \rangle_P$ is monotone deceasing.
For $P \to 0$ the stretch becomes large and $\lambda'(P) \to \infty$. On the other hand for,   $P \to \infty$ the average stretch
turns negative and $\lambda'(P) \to - \infty$. Hence there is a unique pressure, $P_\mathrm{c}$, such that 
$\lambda'(P_\mathrm{c}) = 0$.
 The solution to \eqref{3.16} has exactly two branches, labelled  by $\lambda,-$ for  $0 < P <P_\mathrm{c}$ and 
 $\lambda,+$ for $P_\mathrm{c}< P $. More precisely \eqref{3.17} reads
 \begin{equation}\label{3.17a} 
 P =  \int _\mathbb{R}\mathrm{d}x  \varrho_{\lambda,-}^*(x),\quad 0 < P < P_\mathrm{c},\qquad
  P =  \int _\mathbb{R}\mathrm{d}x  \varrho_{\lambda, +}^*(x), \quad P > P_\mathrm{c},
 \end{equation}
which defines the two branches $P_{\pm}(\lambda)$ with $\lambda <  \lambda(P_\mathrm{c})$. In particular
 \begin{equation}\label{3.17b} 
  \varrho^*(x,P) =\varrho_{\lambda(P),-}^*(x),\quad P < P_\mathrm{c},\qquad 
  \varrho^*(x,P) =\varrho_{\lambda(P),-}^*(x),\quad P > P_\mathrm{c}.
 \end{equation}
 
 Our interest are the Toda GGE averages 
 \begin{equation}\label{3.21} 
\lim_{N\to\infty} \frac{1}{N} \langle Q^{[n],N}\rangle_N = \langle Q^{[n]}_{0}\rangle,\qquad 
\lim_{N\to\infty} \frac{1}{N} \langle Q^{[\mathrm{s}],N}\rangle_N = \langle Q^{[\mathrm{s}]}_0\rangle,
 \end{equation}
where $\langle \cdot\rangle$ stands for the infinite volume GGE average. These expectations  can be computed through first order derivatives of the Toda free energy,
starting from the expression \eqref{3.14}. 
 For the average of the $n$-th conserved field one obtains
 \begin{equation}\label{3.22} 
 \langle Q^{[n]}_{0}\rangle = \partial_\kappa F_\mathrm{toda}(V+\kappa x^n,P)\big|_{\kappa = 0} = \partial_P \partial_\kappa \mathcal{F}_\lambda(\varrho^*(V+\kappa x^n,P))\big|_{\kappa = 0} .
 \end{equation}
 Therefore we first introduce the linearization of $\varrho^*$ as
\begin{equation}\label{3.27} 
\partial_\kappa \varrho^*(V+\kappa x^n,P)\big|_{\kappa = 0} =  \varrho^{*\{n\}}(V,P). 
\end{equation}
Then
\begin{eqnarray}\label{3.23} 
 && \hspace{0pt}  \partial_\kappa \mathcal{F}_\lambda(\varrho^*(V+\kappa x^n,P)\big|_{\kappa = 0} =
\int \mathrm{d}x \varrho^*(x,V,P)x^n + \int \mathrm{d}x V(x) \varrho^{*\{n\}}(x) \nonumber\\
 &&
  - 2 \int_\mathbb{R} \mathrm{d}x\int_\mathbb{R} \mathrm{d}y  \log|x-y| \varrho^{*\{n\}}(x)\varrho^*(y,V,P) + \int_\mathbb{R} \mathrm{d}x 
  \varrho^{*\{n\}}(x) \log \varrho^*(x,V,P),
  \end{eqnarray}
where we used that $\int \mathrm{d}x \varrho^{*\{n\}}(x)= 0$.
 Integrating the Euler-Lagrange equation  \eqref{3.16} at $\lambda = \lambda(P)$ against $\varrho^{*\{n\}}$, the terms on the right of \eqref{3.23} vanish and
 \begin{equation}\label{3.24} 
 \langle Q^{[n]}_{0}\rangle 
 =  \int_\mathbb{R}\mathrm{d}x \partial_P\varrho^*(x,P) x^n. 
\end{equation}
For the average stretch one differentiates \eqref{3.20} to obtain
\begin{equation}\label{3.25} 
\langle Q^{[\mathrm{s}]}_{0}\rangle = \partial_P F_\mathrm{toda}(P) = \lambda'(P) = \Big(\int _\mathbb{R}\mathrm{d}x  \partial_\lambda\varrho_\lambda^*(x) \Big)^{-1} .
\end{equation}

By the same procedure one can work out the susceptibility matrix 
\begin{equation}\label{3.26} 
\mathsf{C}_{m,n} = \partial_\kappa \partial_{\kappa'} F_\mathrm{toda}(V+\kappa x^n +\kappa' x^m,P)\big|_{\kappa,\kappa' = 0}.
\end{equation}
We note that $\varrho^{*\{n\}}$ satisfies the linear equation
\begin{equation}\label{3.28} 
\varrho^{*}(x) x^n  - 2\int_\mathbb{R} \mathrm{d}y \log|x-y| \varrho^{*}(x)\varrho^{*\{n\}}(y) + \varrho^{*\{n\}}(x) = 0.
\end{equation}
Then 
\begin{eqnarray}\label{3.29} 
&&\hspace{-30pt}\partial_\kappa \partial_{\kappa'} F_\mathrm{toda}(V+\kappa x^n +\kappa' x^m,P)\big|_{\kappa,\kappa' = 0} 
\\
 &&\hspace{-10pt}
= \partial_P\Big(  - 2 \int_\mathbb{R} \mathrm{d}x\int_\mathbb{R} \mathrm{d}y  \log|x-y|\varrho^{*\{n\}}(x)\varrho^{*\{m\}}(y)  + \int_\mathbb{R} \mathrm{d}x (\varrho^*(x))^{-1} 
\varrho^{*\{n\}}(x)\varrho^{*\{m\}}(x)\Big).  \nonumber
\end{eqnarray}

We return to the Lax matrix $L_N$ with eigenvalues $\ell_1,...,\ell_N$ and hence with empirical density
\begin{equation}\label{3.29a}
\rho_{L,N}(\mathrm{d}x) = \frac{1}{N} \sum_{j=1}^N \delta (x- \ell_j)\mathrm{d}x.
\end{equation}
From \eqref{3.24} one deduces that on average
\begin{equation}\label{3.29b}
\lim_{N\to\infty} \rho_{L,N}(\mathrm{d}x) = \rho_{L}(x) \mathrm{d}x =  \partial_P\varrho^*(x,P) \mathrm{d}x.
\end{equation}
As to be discussed in Sect. 4, when integrated against some smooth test function, the con\-vergence is actually almost surely.
Thus $\partial_P\varrho^*(x,P)$ is the deterministic density of states (DOS) of the Lax matrix, whose matrix elements are  distributed according to the GGE at parameters $V,P$.
In particular, $\partial_P\varrho^*(x,P) \geq 0$, which cannot be deduced directly from its definition.
On the hydrodynamic scale, by assumption, the (deterministic) local DOS is initially changing slowly in space and, because of the conservation law, this feature is expected to persist as slow variation in space-time. However $\partial_P\varrho^*(x,P)$ does not carry any information about the average stretch
$\lambda'(P)$. Another option, comprehending all data, would be the two branches of
$\partial_\lambda\varrho_{\lambda, \pm}^*$. The corresponding  
pressure is defined through \eqref{3.17a}. Since
\begin{equation}\label{3.30} 
\partial_\lambda \varrho_{\lambda,\pm}^*(x) = P_\pm'(\lambda)  \partial_P \varrho^*(x,P), 
\end{equation}
``$-$'' referring to $0 < P < P_\mathrm{c}$ and ``$+$'' referring to $P_\mathrm{c} < P$, one concludes that 
\begin{equation}\label{3.31} 
\int_\mathbb{R} \mathrm{d}x\partial_\lambda \varrho_{\lambda,\pm}^*(x) = \frac{1}{\langle r_0 \rangle_{P}},\quad
\langle r_0 \rangle_P \int_\mathbb{R} \mathrm{d}x\partial_\lambda \varrho_{\lambda,\pm}^*(x)x^n =  \langle Q^{[n]}_{0}\rangle
\end{equation}
with $\langle r_0 \rangle_P \partial_\lambda \varrho_{\lambda,\pm}^*(x) \geq 0$ and normalized to 1.

Equation \eqref{3.20}, and its consequences \eqref{3.24}, \eqref{3.25}, \eqref{3.29b} are the \textbf{main result} of our contribution. For the Toda chain, they establish exact expressions 
for the GGE free energy and the GGE average of the
conserved fields. 

%%%%%%%%%%%%%%%%%%%%%%%%%%%%%%%%%%%%%%%%%%%%%%%%%%%%%%%%%%%%%%%
\section{Mean-field limit for Dyson's Brownian motion}
\label{sec4}
\setcounter{equation}{0} 

While in the context of the Toda lattice only the static mean-field limit is of relevance, the dynamic approach to this limit turns out to be instructive. 
In particular, additional identities will be generated.  We thus
 consider the stochastic particle system on $\mathbb{R}$ governed by
\begin{equation}\label{4.1}
dx_j(t) = -V'(x_j(t))dt + \frac{1}{N}\sum_{i = 1,i\neq j}^N \frac{2\alpha}{x_j(t) - x_i(t)} dt + \sqrt{2} db_j(t), \quad j = 1,...,N, \quad\alpha \geq 0 , 
\end{equation}
with $\{b_j(t), j = 1,...,N\}$ a collection of independent standard Brownian motions.  This is Dyson's Brownian motion in an external  potential $V$. 
For easier comparison we have chosen the more neutral $\alpha$, instead of $P$.
The interaction has strength $1/N$, which corresponds to a  standard mean-field limit. We assume an initial measure such that 
for all smooth test functions, $f$, the initial empirical density has the almost sure limit
$\rho_0$,
 \begin{equation}\label{4.2}
\lim_{N \to \infty} \frac{1}{N} \sum_{j = 1}^N f(x_j) = \int_\mathbb{R} \mathrm{d}x\rho_0(x) f(x). 
\end{equation} 
Then also for $t>0$  
\begin{equation}\label{4.3}
\lim_{N \to \infty} \frac{1}{N} \sum_{j = 1}^N f(x_j(t)) = \int_\mathbb{R} \mathrm{d}x \rho(x,t) f(x) 
\end{equation} 
almost surely, where $\rho(x,t)$ is the solution to the mean-field equation
\begin{equation}\label{4.4}
\partial_t \rho(x,t) = \partial_x\Big( V'(x)\rho(x,t) - 2\alpha \int _\mathbb{R}\mathrm{d}y\frac{1}{x-y}\rho(y,t) \rho(x,t)  + \partial_x  \rho(x,t)\Big)
\end{equation} 
with initial condition $\rho(x,0) = \rho_0(x)$. 

The unique stationary probability measure of \eqref{4.1} is given by $\mu_{\alpha,N}^\mathrm{MF}$ with density 
\begin{equation}\label{4.4a}
(Z_{\alpha,V,N})^{-1} \exp\Big[- \sum_{j=1}^N V(x_j) 
+\frac{\alpha}{N}\sum_{i \neq j=1}^N  \log| x_i - x_j| \Big].
\end{equation} 
The corresponding  empirical  density converges to 
the stationary solution of \eqref{4.4}. More precisely, under $\mu_{\alpha,N}^\mathrm{MF}$, for all smooth test functions $f$, 
\begin{equation}\label{4.5}
\lim_{N \to \infty} \frac{1}{N} \sum_{j = 1}^N f(x_j) = \int_\mathbb{R}\mathrm{d}x \rho_\mathrm{s}(x) f(x)
\end{equation}
 almost surely, where  $\rho_\mathrm{s}$ satisfies  
 \begin{equation}\label{4.7}
\int _\mathbb{R}\mathrm{d}x\rho_\mathrm{s}(x) (\partial_x-V'(x))\partial_x f(x) + \alpha\int _\mathbb{R}\mathrm{d}x\int _\mathbb{R}\mathrm{d}y \rho_\mathrm{s}(x)\rho_\mathrm{s}(y)
\frac{\partial_x f(x) - \partial_yf(y)}{x-y}=0.
\end{equation}
Point-wise this equation reads
\begin{equation}\label{4.6}
\partial_x\Big( V'(x)\rho_\mathrm{s}(x) - 2\alpha\int _\mathbb{R}\mathrm{d}y\frac{1}{x-y}\rho_\mathrm{s}(y) \rho_\mathrm{s}(x)  + \partial_x  \rho_\mathrm{s}(x)\Big) = 0,\quad
  \int _\mathbb{R}\mathrm{d}x\rho_\mathrm{s}(x) = 1. 
\end{equation} 

Since Dyson's Brownian motion is time-reversible, stationarity is equally defined by the large round bracket in \eqref{4.6} to vanish. The respective free energy (large deviation) functional is \eqref{3.7}, setting $P = \alpha$. Its minimizer $\rho^*$ satisfies a corresponding Euler-Lagrange equation.
Differentiating this equation with respect to $x$ and  then multiplying by $\rho^*$ yields
\begin{equation}\label{4.8} 
  V'(x)\rho{^*}(x) -  2 \alpha \int \mathrm{d}y \frac{1}{x-y} \rho^*(x)\rho^*(y) + \partial_x\rho^*(x) = 0.
 \end{equation}
Thus we conclude $\rho_\mathrm{s} = \rho^*$.

Under general conditions on $V$, the law of large numbers \eqref{4.3}, \eqref{4.4} has been proved by C\'{e}pa and L\'{e}pingle  \cite{CL97}. 
Much studied is also the strong coupling regime \cite{I01}, for which  $2\alpha/N$ in \eqref{4.1} is substituted by $\beta >0$. The two regimes have a very different $N$-dependence. In mean-field the free energy 
is of order $N$, the empirical density is strictly positive with size of order 1, and the fluctuations are of order $\sqrt{N}$. On the other side, for fixed $\beta$ there is a strong repulsion between particles. Thus the free energy is of order $N^2$ and the empirical density is extended over a region of size $\sqrt{N}$, while the shape function 
is generically of compact support. Hence the diffusion term is of order $1/N$ and in the variational formula the entropy term may be dropped. Particles can hardly move any more and fluctuations are of order 1. However the large $\alpha$ limit of  $\rho_\mathrm{s}$ seems to match with the strong coupling 
steady state in the limit of small $\beta$.

 %%%%%%%%%%%%%%%%%%%%%%%%%%%%%%%%%%%%%%%%
%%%%%%%%%%%%%%%%%%%%%%%%%%%%%%%%%%%%%%%%%%
\section{Toda chain in thermal equilibrium}
\label{sec5}
\setcounter{equation}{0} 

Thermal equilibrium corresponds to the special choice $V(x) = \tfrac{1}{2}x^2$,  in which case 
 the stationary density $\rho_\mathrm{s}$ is known explicitly  and given by 
\begin{equation}\label{5.1}
\rho_\mathrm{s}(x) = (2\pi)^{-\frac{1}{2}} \mathrm{e}^{-\frac{1}{2}x^2} |\hat{f}_\alpha(x)|^{-2},  
\end{equation} 
where
\begin{equation}\label{5.2}
\hat{f}_\alpha(x) = \int_0^\infty f_\alpha(t) \mathrm{e}^{\mathrm{i}xt}\mathrm{d}t, \quad f_\alpha(t) = (\alpha/\Gamma(\alpha))^\frac{1}{2}t^{\alpha - 1}  \mathrm{e}^{-\frac{1}{2}t^2}. 
\end{equation}
Physically, one should introduce the proper inverse temperature and pressure. This will be  done at the end of this section. 
 
Opper  \cite{Opper} derived \eqref{3.16} and \eqref{3.17} for $V(x) = \tfrac{1}{2}x^2$ through the semi-classical limit of the quantum Toda chain and obtained the exact solution  \eqref{5.1},  \eqref{5.2},  see also the earlier work \cite{T84} and the discussion in \cite{T89}, Section 6.6.
 Of course, the connection to random matrices was not known yet. The more recent investigation \cite{ABG12} starts from a modified
 Dyson's Brownian motion, such that in a particular limit one arrives at the mean-field interaction. As a common feature  of these studies 
the Stieltjes transform of \eqref{4.6} is shown to satisfy a local quadratic equation. It can be linearized to arrive at 
a linear  second order differential equation, the Weber equation, which then leads to the explicit solution   \eqref{5.1}, \eqref{5.2}.
 For small $\alpha$, $\rho_\mathrm{s}$ is close to the unit Gaussian, while $\rho_\mathrm{s}$ converges to the Wigner semi-circle law in the limit $\alpha \to \infty$. 
 A related study has been carried out for $\beta$-Wishart ensembles \cite{ABM12}.
 
 Note that $\frac{d}{d\alpha}\big(\alpha \rho_\mathrm{s}\big)$ is the DOS of the two-sided Lax matrix, $L$, defined through \eqref{2.19} in the two-sided limit $N \to \infty$. $L$ can be viewed as a random Jacobi matrix and from this perspective Duy and Shirai \cite{DS15} and Duy 
\cite{D18} independently employ the connection to the Dumitriu and Edelman theorem to establish the exact DOS.
 
 Choosing  $f$  in \eqref{4.7} to be a monomial  one obtains a recursion relation for the moments of $\rho_\mathrm{s}$. The odd moments vanish and for the even moments, 
$c_n = \int _\mathbb{R}\mathrm{d}x\rho_\mathrm{s}(x)x^{2n}$, 
\begin{equation}\label{5.3}
c_n = (2n-1)c_{n-1}  +  \alpha \sum_{j=0}^{n-1}c_{n-1-j} c_j,\quad c_0 = 1.
\end{equation} 
This recursion relation is stated already in \cite{DS15}.

The large $\alpha$ behavior can read off from \eqref{2.5} by expanding $\mathrm{tr}\big[(L_N)^n\big]$.   Each term corresponds to a random walk,
step size $0,\pm1$, starting and ending at site $j$, and weight resulting from the thermal average. For large $\alpha$  to leading order only paths with no step of size $0$ count and hence
\begin{equation}\label{20a}
c_n =  \alpha^n   \frac{1}{n+1}
 \begin{pmatrix}
2n\\
n
\end{pmatrix}. 
\end{equation}  
On the right side one observes the Catalan numbers. Hence $\rho_\mathrm{s} $ is the normalized Wigner semi-circle probability distribution function
\begin{equation}\label{21}
\rho_\mathrm{s}(x) = \frac{1}{2\pi \alpha} \sqrt{4\alpha - x^2}, \quad x^2 \leq 4\alpha,
\end{equation}
and the Lax matrix DOS equals $\big(\pi \sqrt{4\alpha - x^2}\big)^{-1}$ with $x^2 \leq 4\alpha$.

As to be explained in Section \ref{sec6} below, also the Gaussian fluctuations are of interest in connection with hydrodynamics linearized around thermal equilibrium.
The CLT for the Lax matrix is studied in \cite{NT18} with a somewhat different perspective. Our goal is to have reasonably explicit expressions for the covariance matrix.

We start from stationary Dyson's Brownian motion \eqref{4.1} with $V(x) = \tfrac{1}{2}x^2$ and define  the fluctuation field
\begin{equation}\label{5.4}
\phi_N(f,t) = \frac{1}{\sqrt{N}} \sum_{j=1}^N \Big(f(x_j(t)) - \int _\mathbb{R}\mathrm{d}x   \rho_\mathrm{s}(x) f(x)\Big)
\end{equation}
for some  smooth test function $f$. Then, in distribution,
 \begin{equation}\label{5.5}
\lim_{N\to \infty} \phi_N(f,t) = \phi(f,t) 
\end{equation} 
with $\phi(f,t)$ an infinite-dimensional stationary Ornstein-Uhlenbeck process  governed by
\begin{eqnarray}\label{5.6}
&&\hspace{-40pt}\frac{d}{dt} \phi(f,t) = \phi\big((\partial_x - x)\partial_x f,t\big)\nonumber\\&&\hspace{-30pt} + \alpha\int _\mathbb{R}\mathrm{d}x\int _\mathbb{R}\mathrm{d}y
\frac{\partial_x f(x) -\partial_y f(y)}{x-y}\big( \rho_\mathrm{s}(x) \phi(y,t) + \phi(x,t)\rho_\mathrm{s}(y)\big) +\xi(\sqrt{2\rho_\mathrm{s}}\partial_x f,t),
\end{eqnarray} 
where $\xi$ denotes standard space-time white noise. The stationary covariance $C$, with kernel $C(x,y)$, is then defined through the equation
\begin{equation}\label{5.7}
\langle Df,C g\rangle + \langle f,C Dg\rangle = - 2\int _\mathbb{R}\mathrm{d}x \rho_\mathrm{s}(x) \partial_x f(x) \partial_x g(x),
\end{equation} 
where
\begin{equation}\label{5.8}
Df(x) = (\partial_x - x)\partial_x f(x) + 2\alpha \int _\mathbb{R}\mathrm{d}y  \rho_\mathrm{s}(y) \frac{\partial_x f(x) - \partial_yf(y)}{x-y}
 \end{equation} 
and $\langle\cdot;\cdot \rangle$ denotes the usual $L^2$ inner product. A corresponding result for strong coupling regime has been proved in 
\cite{I01}. The martingale identities used there seem to extend to our case.
 
For $\alpha = 0$ one confirms that the solution to \eqref{5.8} is given by
\begin{equation}\label{5.9}
C_0(x,y) = \rho_\mathrm{G}(x) \delta(x -y) - \rho_\mathrm{G}(x)\rho_\mathrm{G}(y)
 \end{equation} 
with $\rho_\mathrm{G}$ the unit Gaussian. For $\alpha >0$ the most instructive choice for the  basis of functions seems to be simply the powers $x^n, n\geq 0$. We set 
\begin{equation}\label{5.10}
C_{m,n} = \int _\mathbb{R}\mathrm{d}x\int _\mathbb{R}\mathrm{d}y C(x,y) x^my^n, \quad m,n\geq 1, \quad C_{m,n}= C_{n,m},
 \end{equation} 
and note the boundary condition $C_{0,n}= 0 = C_{n,0}$ to be used below. By reflection symmetry of \eqref{3.6}, $C_{m,n} =0$ for odd $m+n$. 
Denoting by $\langle \cdot\rangle_{\rho_\mathrm{s}}$ the average with respect to $\rho_\mathrm{s}$ and by  $\rho_\mathrm{s}$ being even, 
for $n>1$ the action of $D$ reads
\begin{equation}\label{5.11}
Dx^n = n\Big( \big((n-1)x^{n-2} -x^n\big) +2\alpha \sum_{j=0,j \,\mathrm{even}}^{n-2}x^{n-2-j} \langle y^{j}\rangle_{\rho_\mathrm{s}}\Big),\quad
Dx = -x,
 \end{equation} 
which defines the matrix $(\Lambda^\mathrm{T})_{m,n}$. In its explicit form 
\begin{eqnarray}\label{5.12}
&&\Lambda = \Lambda^{0} +  \Lambda^{\alpha}, \nonumber\\
&& \Lambda^{0}_{n,n} = - n, \quad n\geq 1, \qquad \Lambda^{0}_{n-2,n} = - n(n-1),  \quad n\geq 2,\nonumber\\
 &&\Lambda^{\alpha}_{n-2-j,n}= 2\alpha n \langle y^{j}\rangle_{\rho_\mathrm{s}}, \quad j = 0,...,n-3,\quad j\,\,\mathrm{even},
 \end{eqnarray} 
while all other matrix elements vanish, $m,n\geq 1$. Then the stationary condition \eqref{5.7} translates into
 \begin{equation}\label{5.13}
(\Lambda^\mathrm{T} C)_{m,n} + ( C\Lambda)_{m,n}= - 2mn\langle x^{m+n -2}\rangle_{\rho_\mathrm{s}}.
 \end{equation} 
 
 One computes $C_{1,1} = \langle x^{0}\rangle_{\rho_\mathrm{s}} =1$. Assume that $C_{m,n}$ is known for $1 \leq m,n \leq \kappa$. Then $C_{m,\kappa +1}$, $m = 1,...,\kappa+1$, 
 can be obtained iteratively through \eqref{5.13} upon starting from the boundary. The remaining matrix elements from the set $1 \leq m,n \leq \kappa +1$ follow by symmetry. Hence, not only the mean, but also the covariance matrix allows for a recursive computational scheme. 
 
 Finally, we extend our results for the Toda lattice at thermal equilibrium with inverse temperature $\beta$ and physical pressure $P$. For this purpose we have to adjust our notation to a more explicit form. We set $\alpha = \beta P$ and add the $\alpha$- and $\beta,P$-dependence in the obvious way as $\rho_{\mathrm{s},\alpha}$, $L(P,\beta)$. 
Let us recall the random Lax matrix 
\begin{equation}\label{5.14}
(L_\alpha)_{j,j} =  \xi_{\mathrm{G},j}, \quad (L_\alpha)_{j,j+1} = (L_\alpha)_{j+1,j} = \tfrac{1}{\sqrt{2}}\chi_{2\alpha,j},
\end{equation} 
$(L_\alpha)_{i,j}=0$ otherwise, $i,j \in \mathbb{Z}$, and all random matrix elements being independent. Now, with $\tilde{H} = H + P\sum_j r_j$,
\begin{equation}\label{5.15}
\mathrm{e}^{-\beta\tilde{H}} =   \mathrm{e}^{-\beta\frac{1}{2} \mathrm{tr}(L_\alpha^2)} = \mathrm{e}^{-\frac{1}{2}\mathrm{tr}(L(P,\beta)^2)},
\end{equation}
which defines the thermal Lax matrix $L(P,\beta)$. In other words,
\begin{equation}\label{5.16}
L(P,\beta) = \frac{1}{\sqrt{\beta}}L_\alpha, \quad \alpha = P\beta,
\end{equation}
in distribution.
Therefore the DOS scales as
\begin{equation}\label{5.17}
\rho_{P,\beta} (x) \mathrm{d}x= \sqrt{\beta} \rho_{\mathrm{s},\alpha} (\sqrt{\beta} x) \mathrm{d}x, 
\end{equation}
which for the moments implies
\begin{equation}\label{5.18}
\langle Q^{[{n}]}_{0} \rangle_{P,\beta} = \beta^{-n/2} \frac{d}{d\alpha}\Big(\alpha \int_\mathbb{R} \mathrm{d}x \rho_{\mathrm{s},\alpha}(x) x^n\Big),
\quad n \geq 1.
\end{equation}
The stretch average is treated separately by setting 
\begin{equation}\label{5.19}
\langle f(r) \rangle_r = \big(\beta^{-\beta P} \Gamma(\beta P)\big)^{-1}\int_\mathbb{R} \mathrm{d}r \exp\big[-\beta (e^{-r} +Pr)\big]f(r)
\end{equation}
and 
\begin{equation}\label{5.20}
\langle Q^{[\mathrm{s}]}_{0} \rangle_{P,\beta} = \langle r \rangle_r.
\end{equation}

%%%%%%%%%%%%%%%%%%%%%%%%%%%%%%%%%%%%%%%%
%%%%%%%%%%%%%%%%%%%%%%%%%%%%%%%%%%%%%%%%%%
\section{Linearized hydrodynamics}
\label{sec6}
\setcounter{equation}{0} 

As discussed in \cite{DS17}, for writing down the linearized Euler equation two matrices are required. In principle the scheme works for any GGE, but in our context we linearize around thermal equilibrium, hence the thermal average $\langle \cdot \rangle_{P,\beta}$ appears with $\beta$ the inverse temperature and $P$ the physical pressure, while
$\beta P$ is the Lagrange multiplyer for $r_j$. The first matrix is the static susceptibility defined by
\begin{equation}\label{6.1}
\mathsf{C}_{m,n} = \sum_{j\in\mathbb{Z}}\langle Q^{[m]}_{j};Q^{[n]}_{0} \rangle_{P,\beta}, \quad m,n = \mathrm{s},1,2,...\,,
\end{equation} 
where $\langle\cdot;\cdot \rangle_{P,\beta}$ stands for the second cumulant. Note that by strict locality the sum consists of finitely many terms only. The second matrix is the cross correlation between fields and currents,
\begin{equation}\label{6.2}
\mathsf{B}_{m,n} = \sum_{j\in\mathbb{Z}}\langle J^{[m]}_{j};Q^{[n]}_{0} \rangle_{P,\beta}, \quad m,n = \mathrm{s},1,2,...\,.
\end{equation} 
 In the  limit of zero lattice spacing, the lattice of labels, the linearized Euler equations read 
\begin{equation}\label{6.3}
\partial_t u_n(x,t) + \sum_{m=\mathrm{s},1}^\infty \mathsf{A}_{n,m} \partial_x u_m(x,t) = 0,\quad n = \mathrm{s},1,..., \quad \mathsf{A} = \mathsf{B}\mathsf{C}^{-1},
\end{equation} 
because of the transformation from intensive to extensive variables. In passing we recall that the matrix $\mathsf{B}\mathsf{C}^{-1}\mathsf{B}$ is the Drude weight of the Toda chain in thermal equilibrium, in the sense that $\big(\mathsf{B}\mathsf{C}^{-1}\mathsf{B}\big)_{m,n}$ equals the long time asymptotics of the correlation between the $m$-th and $n$-th total current  \cite{Zotos,DS17,MS16}. 

Let us deviate for a moment to explain why, despite its asymmetric looking definition, one finds that $\mathsf{B}$ is actually symmetric, see also  \cite{S14}. For this purpose first note that 
for an initial GGE the process $Q^{[n]}_{j}(t)$ is space-time stationary. Hence
 \begin{equation}\label{6.3a}
\langle Q^{[m]}_{j}(t);Q^{[n]}_{0}(0) \rangle_{P,\beta} = \langle Q^{[m]}_{0}(0);Q^{[n]}_{-j}(-t) \rangle_{P,\beta},
\end{equation} 
which upon summing over $j$ and differentiating at $t = 0$ turns into
\begin{equation}\label{6.3b}
\sum_{j \in \mathbb{Z}} j \langle \big(J^{[m]}_{j} - J^{[m]}_{j+1};Q^{[n]}_{0}\big) \rangle_{P,\beta} = 
-\sum_{j \in \mathbb{Z}} j\langle Q^{[m]}_{0}(0); \big(J^{[n]}_{-j} - J^{[n]}_{-j+1} \big)\rangle_{P,\beta}.
\end{equation} 
Thus $\mathsf{B} = \mathsf{B}^\mathrm{T}$ by partial summation.

In connection with the molecular dynamics \cite{KD16}, of particular interest is the field-field time correlation
\begin{equation}\label{6.4}
S_{m,n}(j,t) =  \sum_{j\in\mathbb{Z}}\langle Q^{[m]}_{j}(t);Q^{[n]}_{0}(0) \rangle_{P,\beta}, \quad m,n = \mathrm{s},1,2,...\, .
\end{equation}
For example, up to factors of $2$, the matrix elements $m,n=\mathrm{s},1,2$ would be stretch, momentum, and energy correlations, including their cross-correlations.
In  leading order one would expect that such correlations are equal to the solution to \eqref{6.3} with random initial conditions, whose covariance equals
\begin{equation}\label{6.4a}
\langle u_m(x,0)u(x',0)\rangle = \mathsf{C}_{m,n} \delta (x-x').
\end{equation}
The solution is given by 
\begin{equation}\label{6.5a}
\langle u_m(x,t)u(0,0)\rangle = \mathsf{S}(x,t) = \frac{1}{2\pi}\int_\mathbb{R} \mathrm{d}k \mathrm{e}^{\mathrm{i}kx} \mathrm{e}^{-\mathrm{i} kt\mathsf{A}}\mathsf{C}.
\end{equation}
 Since $ \mathsf{A}\mathsf{C} = \mathsf{B}$,  the more symmetric form reads
\begin{equation}\label{6.6}
\mathsf{S}(x,t) = \frac{1}{2\pi}\int_\mathbb{R} \mathrm{d}k \mathrm{e}^{\mathrm{i}kx} \mathsf{C}^\frac{1}{2}\mathrm{e}^{-\mathrm{i} kt\mathsf{C}^{-\frac{1}{2}}\mathsf{B}\mathsf{C}^{-\frac{1}{2}}}
\mathsf{C}^\frac{1}{2},
\end{equation}
which admits  the spectral resolution 
$\rho_{m,n}(\mathrm{d}\lambda)$ such that
\begin{equation}\label{6.7}
\mathsf{S}_{m,n}(x,t) = \int_\mathbb{R} \rho_{m,n}(\mathrm{d}\lambda) \delta(x -\lambda t), \quad \int_\mathbb{R} \mathrm{d}x\,\mathsf{S}(x,t)
= \mathsf{C} = \int_\mathbb{R} \rho(\mathrm{d}\lambda).
\end{equation}
In particular $\mathsf{S}_{m,n}(x,t)$ scales exactly as $t^{-1}f(x/t)$. On the other hand the true microscopic correlations  attain their scaling behavior only 
for large $j,t$ both of the same order and
\begin{equation}\label{6.5}
S_{m,n}(j,t) \simeq \mathsf{S}_{m,n}(x,t)
\end{equation} 
with $x$ the continuum version of $j$.
Molecular dynamics  simulations of the lowest order equilibrium time correlations are available for parameters $N=1024$, $\beta = 1$, $P=1$ at times 200, 300. The average is over $10^6 - 10^7$ initial thermal samples and
 ballistic scaling is well satisfied \cite{KD16}.

Some general properties of $\mathsf{C},\mathsf{B}$ can be obtained from time-reversal $p_j \leadsto -p_j$ for all $j$, denoted by $\mathcal{R}$.
The equilibrium measure is even under $\mathcal{R}$. By inspection $\mathcal{R} Q^{[n]}_{0} = (-1)^n Q^{[n]}_{0}$ and similarly for the currents 
$\mathcal{R} J^{[n]}_{0} = (-1)^{n+1} J^{[n]}_{0}$. Hence, for all $m,n = \mathrm{s}, 1,2,...\, $, $\mathrm{s}$ considered as even,
\begin{equation}\label{6.8}
\langle Q^{[n]}_{0} \rangle_{P,\beta} = 0,\quad n \,\,\mathrm{odd},\qquad \langle J^{[n]}_{0} \rangle_{P,\beta} = 0,\quad n \,\,\mathrm{even},
\end{equation} 
and
\begin{equation}\label{6.9}
\mathsf{C}_{m,n}  = 0,\quad n+m \,\,\mathrm{odd},\qquad   \mathsf{B}_{m,n} = 0,\quad m+n \,\,\mathrm{even}.
\end{equation} 
We call matrices with the left property in \eqref{6.9} of type $\mathcal{A}_0$ and with  the right property of type $\mathcal{A}_1$. One easily verifies that 
if $C_1,C_2 \in  \mathcal{A}_0$, then $C_1C_2 \in  \mathcal{A}_0$. Also if $C_1 \in  \mathcal{A}_0$ and $B_1 \in  \mathcal{A}_1$, then
$A_1B_1, B_1A_1 \in  \mathcal{A}_0$. Hence  $\mathsf{C}^{-\frac{1}{2}}\mathsf{B}\mathsf{C}^{-\frac{1}{2}} \in  \mathcal{A}_0$. This property ensures 
that $\rho_{m,n}(\mathrm{d}\lambda)= \rho_{m,n}(-\mathrm{d}\lambda)$ and thus $\mathsf{S}(x,t) = \mathsf{S}(-x,t)$.

%%%%%%%%%%%%%%%%%%%%%%%%%%%%%%%%%%%%%%%
%%%%%%%%%%%%%%%%%%%%%%%%%%%%%%%%%%%%%%%
\section{Conclusions}
\label{sec7}
\setcounter{equation}{0} 
Beyond the Toda chain there are several classical integrable models which admit a Lax pair. Thus it would be of interest 
whether  similar ideas as in this note could be used to construct GGEs. For one-dimensional continuum models, as the sinh-Gordon model and the nonlinear Schr\"{o}dinger equation, because of ultraviolet divergencies, the GGE tends to be a singular object. But rather commonly there are lattice versions 
which might be more approachable. Possibly there is some generic structure of which the Toda chain is just one example.

If in the Euler-Lagrange equation \eqref{3.16} one substitutes $\varrho_\lambda^*(x) = \exp[- \varepsilon (x)]$, then
\begin{equation}\label{7.1} 
  V(x) -  2 \int \mathrm{d}y  \log|x-y| \mathrm{e}^{-\varepsilon(y)} - \varepsilon (x)  - \lambda = 0,
 \end{equation}
which can be viewed as the semi-classical limit of the thermodynamic Bethe ansatz (TBA) for the quantum Toda chain.
 In fact, for $ V(x) = \tfrac{1}{2}x^2$, this is how \eqref{7.1} has been derived first \cite{T84,Opper}, see also \cite{GM88}. 
 
 While writing up my results, Benjamin Doyon accomplished a derivation of the Toda GHD \cite{D19}. Doyon starts from the fluid picture of Toda particles moving on the real line, obtains \eqref{7.1} using inverse scattering theory, and thereby confirms that $\partial_\lambda\varrho_{\lambda}^*$ can be interpreted as density of quasi-particles. GHD is written for both, lattice and fluid, coordinate frames. Independently, GHD is derived through a semi-classical limit starting from the quantum Toda chain, also including a discussion of quantum corrections \cite{BCM19}. Both contributions state the same definite prediction for the DOS linked to the currents. Comparing with \eqref{2.13a}, it differs from the Lax matrix DOS only by the extra factor $L_N^\mathrm{off}$, which means that now some information on the eigenvectors of the Lax matrix is required.
In the more recent study \cite{BCS19}, the DOS of the Lax matrix is sampled by Monte-Carlo techniques for quadratic and double well potentials.
 One finds very good agreement with numerically iterating \eqref{4.4} and letting the solution relax to the steady state.  Also, the prediction for the GGE average currents is tested. While an analytical argument is still missing, the simulations are strongly favoring the conjecture. 
  \\\\
\textbf{Acknowledgements}. I am grateful to Duy Trinh and Simone Warzel for most helpful comments at an early stage of this project.
I benefitted from most stimulating exchanges with Benjamin Doyon and thank Christian Mendl for a careful reading of a first draft. When completing my work,
Xiangyu  Cao and Vir Bulchandani informed me about their study of the Toda lattice with domain wall initial conditions. In particular, they pointed out  to me
the two-valuedness  of $P(\lambda)$.
\newpage
\appendix
\section{ Appendix: infinite volume limit, linear ramp}
\label{sec8}
We use Flaschka variables and restate the issue, omitting irrelevant constants. The variables are $\{a_j \in\mathbb{R}_+,b_j \in\mathbb{R}, j = 1,...,N\}$. Then the Toda GGE reads
\begin{equation}\label{8.1} 
\mu_{\mathrm{to},N} = \frac{1}{Z_{\mathrm{to},N}} \prod_{j=1}^N \mathrm{d}a_j  \mathrm{d}b_j(a_j)^{-1 +2P}  \mathrm{e}^{-\mathrm{tr}[V(L_N)]}
\end{equation}
and the linearly ramped Toda measure is given by
\begin{equation}\label{8.2} 
\mu_{\mathrm{ra},N} = \frac{1}{Z_{\mathrm{ra},N}} \prod_{j=1}^N \mathrm{d}a_j  \mathrm{d}b_j(a_j)^{-1 +2(j/N)P} \mathrm{e}^{-\mathrm{tr}[V(L_N)]}.
\end{equation}
Up to inverting the ramp, normalization, and a boundary term, the ramped Toda measure agrees with \eqref{3.1}, \eqref{3.6}.

We will establish three properties.\medskip\\
(i) The following limit exists  and, up to constants, agrees with $F_\mathrm{toda}(P)$,
\begin{equation}\label{8.3} 
- \lim_{N \to \infty} \frac{1}{N} \log Z_{\mathrm{to},N} = F_\mathrm{to}(P).
 \end{equation}
(ii) The infinite volume limit of $\mu_{\mathrm{to},N}$, shifted by $-\tfrac{1}{2}N$, exists. The limit measure, $\mu_\mathrm{to}$, defined on $(\mathbb{R}_+\times \mathbb{R})^\mathbb{Z}$ is stationary under spatial shifts and exponentially mixing.\medskip\\
(iii) For the linearly ramped free energy one obtains
\begin{equation} \label{8.4} 
- \lim_{N \to \infty} \frac{1}{N} \log Z_{\mathrm{ra},N} = F_\mathrm{ra}(P) = \int_0^1 \mathrm{d}u F_\mathrm{to}(uP).
 \end{equation}
 
 Our proof relies on transfer matrix techniques. We set $N = \kappa M$, $M$ integer, and denote $\mathsf{a}_m = (a_{\kappa(m-1)+1},...,a_{\kappa(m-1)+\kappa})$, $\mathsf{b}_m = (b_{\kappa(m-1)+1},...,b_{\kappa(m-1)+\kappa})$,
 $m = 1,...,M$. We break the system into blocks of length $\kappa$. Then the interaction is only between neighboring blocks and,
 denoting by $\mathsf{a},\mathsf{b}$ the variables in block $m$ and  by $\mathsf{a}',\mathsf{b}'$ the variables in block $m+1$,
 one can choose a transfer 
 matrix $\mathcal{T}_P(\mathsf{a},\mathsf{b}|\mathsf{a}',\mathsf{b}')$ such that 
 \begin{equation}\label{8.5} 
Z_{\mathrm{to},N} = \mathrm{tr}_\mathcal{H}\big[(\mathcal{T}_P)^M\big],
\end{equation}
where $\mathrm{tr}_\mathcal{H}$ refers to the trace in the Hilbert space $\mathcal{H} = L^2\big((\mathbb{R}_+\times \mathbb{R})^\tau, \mathrm{d}^\tau\mathsf{a} \mathrm{d}^\tau\mathsf{b}\big)$.
 $\mathcal{T}_P>0$ pointwise, $\mathcal{T}_P$ can be choosen to be symmetric, and $\mathcal{T}_P$ is trace class. By the Perron-Frobenius theorem,
$\mathcal{T}_P$ has thus a non-degenerate, strictly positive eigenvalue, $\lambda_\mathrm{max}(P)$, which is separated by a gap from the rest of the spectrum, and   
 
 \begin{equation}\label{8.6} 
 \lim_{N \to \infty} \frac{1}{N} \log Z_{\mathrm{to},N} =  \frac{1}{\kappa}\log  \lambda_\mathrm{max}(P).
\end{equation}
By an extension of the  argument, expectation values of local observables have an infinite volume limit. Exponential mixing is a consequence of the spectral gap.

For the ramped partition function the transfer matrix becomes now $m$-dependent. By perturbation theory, for the transfer matrix from block $m$  to block $m+1$ the maximal eigenvalue equals 
$\lambda_\mathrm{max}(Pm/M)$ up to an error of order $1/N$. Also the spectral gap is at most shifted  by order $1/N$. Hence, for large $N$, up to an error of order 
$\mathrm{e}^{-\gamma N}$ with the constant $\gamma >0$ related to the spectral gap,  one arrives
at
\begin{equation}\label{8.7} 
  \frac{1}{N} \log Z_{\mathrm{ra},N} =   \frac{1}{N}\sum_{m=1}^M \log  \lambda_\mathrm{max}(Pm/M),
 \end{equation}
which in the limit $N\to \infty$ converges to the right hand side of \eqref{8.4}.

Our proof relies heavily on the assumption \eqref{2.14a} assuring a finite power series. Once having moved to the mean-field limit, any confining potential  with some minimal smoothness would do.

\end{document}